\documentclass[12pt]{article}

\usepackage{textcomp}
\usepackage{chet}

\draftmode 

\preprint{UCSD-PTH-11-08}

\title{The \texttt{chet} package}

\author{Andreas Stergiou$^{\ast,}$\email{stergiou@physics.ucsd.edu} and
Author2$^{\ast,\dagger,}$\email{address2@example.com}}

\affiliation{$^{\ast}$Department of Physics, University of California, San
Diego, La Jolla, CA 92093 USA\\ $^{\dagger}$Department of Physics,
University of Somewhere Else, \ldots}

\abstract{This is a sample produced using \texttt{chet}. This package is
inspired by Paul Ginsparg's \texttt{harvmac}, but uses \LaTeXe\ instead of
\TeX. The commands provided are to be used as faster alternatives to
\LaTeXe's default environments (which can all still be used with
\texttt{chet}).

\begin{center}
  (\texttt{chet} can be found at
  \href{http://www.ctan.org/pkg/chet}{\texttt{http://www.ctan.org/pkg/chet}}.)
\end{center} }

\date{June 2011}

\begin{document}
\maketitle

\toc

\newsec{Basic commands}[SecLabel]
\subsec{Preamble}
To use \verb1chet1 first make sure that the file \verb1chet.sty1 is in your
path. (If you have references you also have to add the file
\verb1chetref.bst1 to your path.) You can then start your document with
\verb1\documentclass{article}1 and include \verb1\usepackage{chet}1 in the
preamble. For labels of equations, sections, etc.\ to
appear on the margins, you can use the command \verb1\draftmode1.

In the preamble of the document one also specifies the preprint number,
authors' email addresses, and the abstract (see usage in this example
file). If you want to have only one footnote with all email addresses
and without footnote marks, then you can use the command \verb1\emails{}1
inside the \verb1\author{}1 environment. If a specific date is desired,
then just include \verb1\date{}1 with the desired date in the preamble of
your document, and the default current date on the bottom left of the title
page will be substituted with the one you specified.

It is suggested that authors compile straight to \texttt{pdf} with
\texttt{pdflatex}, i.e.\ following \TeX\textrightarrow\verb1PDF1. The
compilation method
\TeX\textrightarrow\verb1DVI1\textrightarrow\verb1PS1\textrightarrow\verb1PDF1
is obsolete and redundant, and should not be used. As far as I know the
only problem that arises frequently with \TeX\textrightarrow\verb1PDF1 is
the inability to obtain \verb1psfrag1 replacements in \verb1eps1 figures;
that can be taken care of very easily with the package \verb1pstool1.

\subsec{Sections}
Sections start with the command \verb1\newsec{}[]1. The first argument is
the name of the section, while the second provides the label. You can refer
back to sections simply by putting a slash in front of their label. For
example, if you write \verb1\newsec{Name}[Label]1 you can just type
\verb1\Label1 in the subsequent text and the number of the section will
appear, e.g.\ you can refer  to section \SecLabel. Note that if you are
referring to a label you define in a later line, for example you want to
refer to a later section, then the default \verb1\ref{Label}1 is
needed.\foot{The same holds for all references to equations defined with
the commands outlined in this section.} Note, also, that the second
argument of the command can be omitted altogether, i.e.\ the command
\verb1\newsec{}1 starts a section but does not give it a label. The
commands \verb1\subsec{}[]1 and \verb1\subsubsec{}[]1 that define
subsections and subsubsections respectively, are similarly defined.

\subsec{Equations}
For equations use the command \verb1\eqn{}[]1. Again, inside \verb1{}1 you
write the equation and inside \verb1[]1 the label, if you want one. An
equation number will appear only if you do type \verb1[Label]1.\foot{If you
leave the \texttt{[]} empty, the equation is going to get a number but not
a label. If you don't type the \texttt{[]} at all, then the equation will
have no number.} For example, If you give the label \verb1EqMagic1 to an
equation, \eqn{e^{i\pi}+1=0,}[EqMagic] then you can just type
\verb1\EqMagic1 to reference it, \EqMagic. For aligned equations with one
number in the vertical middle use the command \verb1\eqna{}[]1. A single
\verb1&1 indicates the alignment point, while \verb1\\1 indicates a line
break. For example, \eqna{\cos^2 \theta+\sin^2 \theta &=1, \\ \cos^2
\theta-\sin^2 \theta&=\cos 2\theta.}[EqTrig] You can later refer to
equation \EqTrig with \verb1\EqTrig1.

Commands that simplify the writing of subequations are also supplied for
two, three, and four subequations. They are, respectively,
\verb1\twoseqn{}[]{}[][]1, \verb1\threeseqn{}[]{}[]1 \verb1{}[][]1, and
\verb1\fourseqn{}[]{}[]{}[]{}[][]1.  Each pair of \verb1{}[]1 receives an
equation and a label,\foot{If you don't want to label a subequation leave
the corresponding \texttt{[]} empty.} while the last \verb1[]1 is used for
an overall label and can be omitted. Each of the equations has an \verb1&1
at the alignment point. For example, equation \EqTrig could be written as
\twoseqn{\cos^2 \theta+\sin^2 \theta&=1,}[FTrig]{\cos^2 \theta-\sin^2
\theta&=\cos 2\theta.}[STrig][TrigAll] You can then refer to \FTrig,
\STrig, or \TrigAll. More complicated structures with subequations can be
achieved with the corresponding \texttt{amsmath} environment. Note that all
equation environments define labels that can be used only later in the text
with \verb1\Label1. The original \verb1\eqref{Label}1 is otherwise needed.
In the rare occasion that the name of your label coincides with the name of
a \LaTeXe\ command, you will get an error and the file won't compile. In
that case, just change the name of your label.

\subsec{Citations}
To cite a paper use the command \verb1\rcite{}1. (The default command
\verb1\cite{}1 can still be used.) The syntax is exactly the same as in
\verb1\cite1, but, if \verb1\draftmode1 is used, \verb1\rcite1 presents the
label of the citation as an exponent to the citation number wherever that
appears (except in the bibliography, where the label appears on the left
margin).

The \verb1.bib1 file can be included in the main \verb1.tex1 file,
preferably at the end, right before the \verb1\end{document}1. The way to
do this is with the environment
\begin{verbatim}

\end{verbatim}
Here, \verb1bibname.bib1 should be substituted with the name of the
\verb1.bib1 file that you call in the command
\verb1\bibliography{}1.\foot{The functions supported from the \texttt{.bst}
style file are \texttt{@article}, \texttt{@book}, \texttt{@inbook}, and
\texttt{@inproceedings}.} (See usage in this example file.) Note that you
still have to run Bib\TeX\ to compile the bibliography.

For example, scale does not imply conformal invariance in $4-\epsilon$
dimensions as demonstrated in \rcite{FGS}.

\ack{I would like to thank Ken Intriligator, Ben Grinstein, and
Jean-Fran\c{c}ois Fortin for adopting \texttt{chet}, and for their numerous
suggestions and tips that helped me improve the package.

Several \LaTeX\ packages are called by default by \texttt{chet}. In
alphabetical order, they are \texttt{amsmath}, \texttt{caption},
\texttt{cite}, \texttt{collref}, \texttt{datetime}, \texttt{filecontents},
\texttt{footmisc}, \texttt{geometry}, \texttt{hyperref}, \texttt{manyfoot},
\texttt{pifont}, \texttt{showkeys}, \texttt{tocloft}, \texttt{xparse}, and
\texttt{xspace}. I would like to thank the authors of these great packages
for their amazing work!}

\appendix

\newsec{Other commands}
Commands often used include
\begin{itemize}
    \item  \verb1\toc1: produces the table of contents.
    \item \verb1\foot{}1: produces footnotes.
    \item \verb1\ack{}1: used for acknowledgements.
    \item \verb1\appendix1: used before the appendices.
    \item \verb1\bibliography{}1: produces the bibliography.
\end{itemize}

\newsec{Conference proceedings}
If you want to use the macros for equations and sections defined in
\texttt{chet}, but have to use another \texttt{.sty} file, for example for
conference proceedings, then you can just use the option
\texttt{macrosonly} when you call \texttt{chet}, i.e.\ instead of
\verb1\usepackage{chet}1, include \verb1\usepackage[macrosonly]{chet}1 in
the preamble of your \texttt{.tex} file.

\bibliography{chetdocbib}

\end{document}